\documentclass[12pt]{revtex4}
\usepackage{graphicx}
\usepackage{dcolumn}
\usepackage{bm}

\begin{document}

\noindent {\Large{\bf Entanglement dynamics for the double
Tavis-Cummings model}} \vskip0.5cm \noindent{Zhong-Xiao Man$^{1}$,
Yun-Jie Xia$^{1}$, and Nguyen Ba An$^{2,3}$} \vskip 0.2cm \noindent
{\normalsize $^1$College of Physics and Engineering, Qufu Normal
University, Qufu 273165, China}

\noindent {\normalsize $^2$Institute of Physics and Electronics, 10
Dao Tan, Thu Le, Ba Dinh, Hanoi, Vietnam}

\noindent {\normalsize $^3$School of Computational Sciences, Korea
Institute for Advanced Study, 207-43 Cheongryangni 2-dong,
Dongdaemun-gu, Seoul 130-722, Korea} \vskip 0.2cm \noindent

\vskip 0.6cm
\begin{flushright}
\begin{minipage}{390pt}
\noindent {\bf Abstract.}  A double Tavis-Cummings model (DTCM) is
developed to simulate the entanglement dynamics of realistic quantum
information processing where two entangled atom-pairs $AB$ and $CD$
are distributed in such a way that atoms $AC$ are embedded in a
cavity $a$ while $BD$ are located in another remote cavity $b$. The
evolutions of different types of initially shared entanglement of
atoms are studied under various initial states of cavity fields. The
results obtained in the DTCM are compared with that obtained in the
double Jaynes-Cummings model (DJCM) [J. Phys. B \textbf{40}, S45
(2007)] and an interaction strength theory is proposed to explain
the parameter domain in which the so-called entanglement sudden
death occurs for both the DTCM and DJCM.\vskip 0.2cm


\noindent {\bf QICS.} 03.30.+e Entangling power of quantum
evolutions

\end{minipage}
\end{flushright}

\vskip 0.8cm \noindent {\bf 1 Introduction}\vskip 0.5cm

\noindent Entanglement is not only a key concept to distinguish
between the quantum and the classical worlds, but has also been
viewed as an indispensable resource to perform various intriguing
global tasks in quantum computing and quantum information processing
[1]. However, a notable characteristic of entanglement is its
fragility in practical applications due to unavoidable interaction
with the environment. It is therefore of increasing importance to
understand entanglement from its dynamical behaviors in realistic
systems. As a rule for a global task, entanglement should be shared
between different remote parties who participate in the task. There
are cases like teleportation [2], remote state preparation [3]%
, etc., in which each particle of a multipartite entangled state is
distributed to a separate location. There are also cases in which
the entangled particles should be distributed so that each location
contains several particles. For example, in quantum secret
communication protocol between Alice and Bob [4], an ordered $N$
Einstein-Podolky-Rosen (EPR) pairs are to be shared in such a way
that Alice and Bob each holds one half of the pairs. That is, at
Alice's location there are $N$ particles which interact with one
environment while the other $N$ partner-particles at Bob's location
collectively interact with another environment. This scenario
results in two independent local environments but each of them is
common for one half of the $N$ EPR pairs. A natural question arises
as to how such kind of particle-environment interactions degrade the
originally prepared global entanglement. This question is of
fundamental interest because any quantum protocol depends
essentially on the quality of the shared entanglement. As a first
step to the problem, in this paper, we consider the case of $N=2$
with two pairs of entangled two-level atoms $AB$ and $CD$ prepared
in one of the two types of Bell-like states, namely,
\begin{equation}
|\psi (0)\rangle _{IJ}=\cos (\alpha )|10\rangle _{IJ}+\sin (\alpha
)|01\rangle _{IJ},  \label{b1}
\end{equation}
and
\begin{equation}
|\varphi (0)\rangle _{IJ}=\cos (\alpha )|11\rangle _{IJ}+\sin (\alpha
)|00\rangle _{IJ},  \label{b2}
\end{equation}
where $IJ\in \{AB,CD\}$ and $|0\rangle $ $(|1\rangle )$ is the atomic ground
(excited) state.

For the simplest case of $N=1,$ i.e., either state (\ref{b1}) or
state (\ref {b2}) is concerned for the initial state of a single
atom-pair, the so-called double Jaynes-Cummings model (DJCM) [5-12]
has been extensively adopted to study this problem because it yields
exact analytical results. In the DJCM, each of two entangled atoms
is embedded in an independent cavity and locally interacts with it.
The results obtained within the DJCM for the initial empty cavities
are that for any value of $\alpha $ state (\ref{b1}) loses its
entanglement only at discrete time moments $t_{l}=(l+1/2)\pi /g$
with $l=0,1,2,...$ and $g$ the atom-cavity coupling constant, but
for a certain domain of $\alpha $ state (\ref{b2}) may become
separable at times smaller than $t_{l}$ and remains unentangled for
some duration of time [6]. The latter phenomenon is referred to in
the current literatures as entanglement sudden death (ESD) [13],
which has been experimentally observed in [14,15]. An entangled
state with ESD in evolution is less robust than states without it,
since ESD puts a limitation on the application time of entanglement.
Therefore, studying ESD, especially conditions and parameter domains
for its occurrence, is important from both theoretical and practical
points of view. In Ref. [10] the DJCM is considered again and it is
found that if the cavity fields are initially in Fock states with
nonzero photon numbers then both atomic states $\left| \psi
(0)\right\rangle $ and $\left| \varphi (0)\right\rangle $ would
suffer from ESD for all values of $\alpha .$ The DJCM was also
investigated from other perspectives and it was shown that the
entanglement evolution of atoms is closely related to their energy
variation [9] and there is a natural entanglement invariant
demonstrating the entanglement transfer among all the system's
degrees of freedom [7].

\begin{figure}[tbp]
\centerline{\scalebox{0.8}{\includegraphics{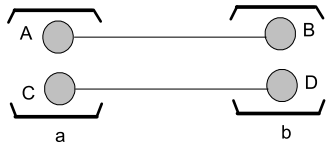}}}
\caption{Schematic representation of two entangled atom-pairs $AB$
and $CD$ of which atoms $A$ and $C$ are located in cavity $a$ but
atoms $B$ and $D $ in another cavity $b$.} \label{fig1}
\end{figure}

For the case of $N=2$ involving two pairs of entangled atoms, the
situation would become more complex than that of $N=1$, because in
each local environment there are two atoms simultaneously
interacting with it. When there are many atoms interacting
resonantly with a single-mode quantized radiation field of one and
the same cavity the exact solution can be obtained by means of the
so-called Tavis-Cummings model (TCM) [16]. Such a single TCM was
used in Refs. [17] and [18] to study entanglement dynamics of two
atoms that are initially prepared in a separable and entangled
state, respectively. In this work we develop the
so-called double Tavis-Cummings model (DTCM) including four two-level atoms $%
A,B,C,D$ and two separate single-mode cavities $a,b$ (see FIG. 1), which
suffices for our purpose to study the entanglement dynamics for case of $N=2$%
. In the DTCM, atoms $A$ $(C)$ and $B$ $(D)$ are initially prepared either
in state (\ref{b1}) or (\ref{b2}), but atoms $A$ and $C$ $(B$ and $D)$ are
located in cavity $a$ $(b)$ and interact with the cavity through the
Tavis-Cummings Hamiltonian. We study the entanglement dynamics of atom-pairs
$AB,$ $CD,$ $AC$ and $BD$ by means of concurrence in dependence on the
initial entanglement type of the atoms and on the initial state of cavity
fields. We compare our results obtained in the DTCM with those obtained in
the DJCM and present an interaction strength theory to explain the parameter
domain in which the atom-pair exhibit ESD for both the DTCM and the DJCM.

Our paper is organized as follows. In Sec. 2 we describe the DTCM
and derive the exact analytical expression for the reduced density
matrix of the atomic subsystem. Section 3 presents detailed analysis
of atomic entanglement dynamics when the initial atom-pairs are
prepared either in state (\ref{b1}) or state (\ref{b2}) and the
initial cavity fields are prepared either in the vacuum state, Fock
state with a non-zero photon number or the thermal state. Finally,
we conclude in Sec. 4.

\vskip 0.8cm \noindent {\bf 2 The double Tavis-Cummings model}\vskip
0.5cm

\noindent The total Hamiltonian of the system of four atoms
$A,B,C,D$ and two cavities $a,b$ (see FIG. 1) in the DTCM can be
written as a sum of two isolated Tavis-Cummings Hamiltonians
\begin{equation}
H=H_{ACa}+H_{BDb},  \label{H}
\end{equation}
with
\begin{equation}
H_{ACa}=\frac{\omega _{0}}{2}(\sigma _{A}^{z}+\sigma _{C}^{z})+\omega
a^{+}a+g\sum_{i=A,C}(a\sigma _{i}^{+}+a^{+}\sigma _{i}^{-}),  \label{H1}
\end{equation}
and
\begin{equation}
H_{BDb}=\frac{\omega _{0}}{2}(\sigma _{B}^{z}+\sigma _{D}^{z})+\omega
b^{+}b+g\sum_{i=B,D}(b\sigma _{i}^{+}+b^{+}\sigma _{i}^{-}),  \label{H2}
\end{equation}
where $\omega _{0}$ $(\omega )$ is the frequency of the atom (cavity
field mode), $a$ $(a^{+})$ is the annihilation (creation) operator
of the field in cavity $a,$ $b$ $(b^{+})$ is the annihilation
(creation) operator of the field in cavity $b,$ $\sigma
_{i}^{+}=\left| 1\right\rangle _{ii}\left\langle 0\right| $ $(\sigma
_{i}^{-}=\left| 0\right\rangle _{ii}\left\langle 1\right| )$ is the
rising (lowering) operator for the transition of atom $i$ and $g$ is
the atom-cavity field coupling constant. Here, we are interested in
the resonant case with $\omega _{0}=\omega $ [16]. The initial
cavity fields are assumed to be either in the vacuum state, the Fock
state with a non-zero photon number or the thermal state. The
general thermal field with its mean photon number $\overline{n}$ is
a weighted mixture of Fock states whose density operator $\rho _{F}$
can be represented as
\begin{equation}
\rho _{F}=\sum_{n=0}^{\infty }P_{n}\left| n\right\rangle \left\langle
n\right| ,  \label{roE}
\end{equation}
with $\left| n\right\rangle $ the Fock state of $n$ photons and $P_{n}$ is
given by
\begin{equation}
P_{n}=\frac{\overline{n}^{n}}{(1+\overline{n})^{n+1}}.  \label{pn}
\end{equation}
By virtue of the general thermal field defined above, through setting $%
P_{n}=\delta _{nl}$ in Eq. (\ref{roE}), we can also study the vacuum state $%
(l=0)$ as well as any Fock states $(l>0)$ of the fields. As for the initial
states of atom-pairs $AB$ and $CD$, we assume both of them to be either in
state (\ref{b1}) or state (\ref{b2}). At $t=0$ the total state involving the
four atoms and two cavities reads
\begin{equation}
\rho (0)=\sum_{i,j,k,l=0}^{1}\sum_{m,n=0}^{\infty }\alpha _{i}\alpha
_{j}\alpha _{k}\alpha _{l}P_{m}^{a}P_{n}^{b}\left| ik,m\right\rangle
_{ACaACa}\left\langle jl,m\right| \otimes \left| i^{\prime }k^{\prime
},n\right\rangle _{BDbBDb}\left\langle j^{\prime }l^{\prime },n\right| ,
\label{initot}
\end{equation}
where $\alpha _{0}\equiv \sin \alpha ,$ $\alpha _{1}\equiv \cos
\alpha $ and $i^{\prime }(j^{\prime },k^{\prime },l^{\prime })\equiv
i(j,k,l)\oplus 1$ (with $\oplus $ an addition mod 2) for state
(\ref{b1}), while $i^{\prime }(j^{\prime },k^{\prime },l^{\prime
})\equiv i(j,k,l)$ for state (\ref{b2}). The evolution operator
$U_{ACa(BDb)}(t)=\exp (-iH_{ACa(BDb)}t)$ for the local interaction
of atoms $AC$ $(BD)$ with cavity $a$ $(b)$ was derived exactly in
Ref. [17]. At any time $t>0$ the state $\rho (0)$ evolves into $\rho
(t)=U_{ACa}(t)U_{BDb}(t)\rho (0)U_{ACa}^{+}(t)U_{BDb}^{+}(t)$ which
can be represented as
\begin{eqnarray}
\rho (t) &=&\sum_{i,j,k,l=0}^{1}\sum_{m,n=0}^{\infty }\alpha _{i}\alpha
_{j}\alpha _{k}\alpha _{l}P_{m}^{a}P_{n}^{b}  \nonumber  \label{fintot} \\
&&U_{ACa}(t)\left| ik,m\right\rangle _{ACaACa}\left\langle jl,m\right|
U_{ACa}^{+}(t)  \nonumber \\
&&\otimes U_{BDb}(t)\left| i^{\prime }k^{\prime },n\right\rangle
_{BDbBDb}\left\langle j^{\prime }l^{\prime },n\right| U_{BDb}^{+}.
\end{eqnarray}
Using the analytical expression of $U_{ACa(BDb)}(t)$ in [17] we have
for $U_{ACa}\left| ik,m\right\rangle _{ACa}$ (similarly for
$U_{BDb}\left|
i^{\prime }k^{\prime },n\right\rangle _{BDb}):$%
\begin{equation}
U_{ACa}(t)\left| ik,m\right\rangle _{ACa}=\sum_{p,q=0}^{1}X_{ik,pq}(m,\tau
)\left| i\oplus p,k\oplus q\right\rangle _{AC}\left|
m-(-1)^{i}p-(-1)^{k}q\right\rangle _{a}  \label{UACa}
\end{equation}
where the functions $X_{ik,pq}(m,\tau )$ with $\tau =gt$ are given in
Appendix A for various possible $i,k,p,q.$ These functions satisfy the
normalization condition
\begin{equation}
\sum_{p,q=0}^{1}|X_{ik,pq}(m,\tau )|^{2}=1
\end{equation}
for any $i,k,m$ and $\tau .$

The reduced density matrix $\rho ^{ABCD}(t)$ of the atomic subsystem can be
obtained by tracing out $\rho (t)$ over the cavity fields, i.e.
\begin{equation}
\rho ^{ABCD}(t)=\text{Tr}_{ab}\rho (t)=\sum_{i,j,k,l=0}^{1}\alpha _{i}\alpha
_{j}\alpha _{k}\alpha _{l}\mathcal{E}_{AC}^{a}\left( \left| ik\right\rangle
_{ACAC}\left\langle jl\right| \right) \otimes \mathcal{E}_{BD}^{b}\left(
\left| i^{\prime }k^{\prime }\right\rangle _{BDBD}\left\langle j^{\prime
}l^{\prime }\right| \right)  \label{rABCD}
\end{equation}
where $\mathcal{E}_{XY}^{c}\left( \left| ik\right\rangle _{XYXY}\left\langle
jl\right| \right) ,$ with $XYc=ACa$ or $BDb,$ represents the map
\begin{eqnarray}
\mathcal{E}_{XY}^{c}\left( \left| ik\right\rangle _{XYXY}\left\langle
jl\right| \right) &\equiv &\sum_{m,m^{\prime }=0}^{\infty
}P_{m}^{c}\left\langle m^{\prime }\right| U_{XYc}(t)\left| ik,m\right\rangle
_{XYcXYc}\left\langle jl,m\right| U_{XYc}^{+}(t)\left| m^{\prime
}\right\rangle  \nonumber \\
&=&\sum_{m=0}^{\infty }\sum_{r,s,u,v=0}^{1}P_{m}^{c}\delta
_{(-1)^{i}r-(-1)^{k}s,(-1)^{j}u-(-1)^{l}v}  \nonumber \\
&&\times X_{ik,rs}(m,\tau )X_{jl,uv}^{*}(m,\tau )\left| i\oplus r,k\oplus
s\right\rangle _{XYXY}\left\langle j\oplus u,l\oplus v\right| .  \label{E}
\end{eqnarray}
The explicit expressions of $\mathcal{E}_{XY}^{c}\left( \left|
ik\right\rangle _{XYXY}\left\langle jl\right| \right) $ are given in
Appendix B for various possible $i,k,j,l.$

\vskip 0.8cm \noindent {\bf 3 Atomic entanglement dynamics}\vskip
0.5cm

\noindent With the formulae derived in the previous section we are
now in the position to analyze the entanglement dynamics of any
atom-pair. By using Eq. (\ref {rABCD}) we can readily get the
reduced density matrix of any pair of atoms by tracing out $\rho
^{ABCD}(t)$ over the degrees of freedom of the remaining atoms. In
two-qubit domains, there exist a number of good measures of
entanglement such as concurrence [19] and negativity [20]. Although
the various entanglement measures may be somewhat different
quantitatively [6], they are qualitatively equivalent to each other
in the sense that all of them are equal to zero for unentangled
states. Here we adopt Wootters' concurrence [19] because of its
convenience in definition, normalization and calculation. The
concurrence $C$ for any (reduced) density matrix $\rho $ of two
qubits is defined as
\begin{equation}
C(\rho )=\max \{0,\sqrt{\lambda _{1}}-\sqrt{\lambda _{2}}-\sqrt{\lambda _{3}}%
-\sqrt{\lambda _{4}}\},  \label{C}
\end{equation}
where $\lambda _{i}$ ($\lambda _{1}\geq \lambda _{2}\geq \lambda _{3}\geq
\lambda _{4})$ are the eigenvalues of the matrix $\zeta =\rho (\sigma
_{y}\otimes \sigma _{y})\rho ^{*}(\sigma _{y}\otimes \sigma _{y}),$ with $%
\sigma _{y}$ a Pauli matrix and $\rho ^{*}$ the complex conjugation
of $\rho $ in the standard basis. For separate states $C(\rho )=0,$
whereas for maximally entangled states $C(\rho )=1.$ In particular,
if $\rho $ is of the X-form [21],
\begin{equation}
\rho ^{IJ}=\left(
\begin{array}{cccc}
\varrho _{11}^{IJ} & 0 & 0 & \varrho _{14}^{IJ} \\
0 & \varrho _{22}^{IJ} & \varrho _{23}^{IJ} & 0 \\
0 & \varrho _{32}^{IJ} & \varrho _{33}^{IJ} & 0 \\
\varrho _{41}^{IJ} & 0 & 0 & \varrho _{44}^{IJ}
\end{array}
\right) ,
\end{equation}
where $\varrho _{kk}^{IJ}$ are real positive and $\varrho _{kl}^{IJ}=\left(
\varrho _{lk}^{IJ}\right) ^{*}$ are generally complex, then the concurrence (%
\ref{C}) simplifies to
\begin{equation}
C^{IJ}=2\max \{0,|\varrho _{23}^{IJ}|-\sqrt{\varrho _{11}^{IJ}\varrho
_{44}^{IJ}},|\varrho _{14}^{IJ}|-\sqrt{\varrho _{22}^{IJ}\varrho _{33}^{IJ}}%
\}.  \label{CX}
\end{equation}
Since both states (\ref{b1}) and (\ref{b2}) of the atoms take on and
preserve the X-form in their evolution, Eq. (\ref{CX}) is very useful
throughout this work.

\vskip 0.8cm \noindent {\bf 3.1 $\left| \psi (0)\right\rangle $ type
initial state for atom-pairs $AB$ and $CD$}\vskip 0.5cm

We first consider the case when both the atom-pairs $AB$ and $CD$
are initially prepared in state (\ref{b1}). In accordance with Eq.
(\ref{rABCD}) the reduced density matrix of the atomic subsystem at
any time $t$ is
\begin{equation}
\rho _{I}^{ABCD}(t)=\sum_{i,j,k,l=0}^{1}\alpha _{i}\alpha _{j}\alpha
_{k}\alpha _{l}\mathcal{E}_{AC}^{a}\left( \left| i,k\right\rangle
_{ACAC}\left\langle j,l\right| \right) \otimes \mathcal{E}_{BD}^{b}\left(
\left| i\oplus 1,k\oplus 1\right\rangle _{BDBD}\left\langle j\oplus
1,l\oplus 1\right| \right) ,  \label{rI}
\end{equation}
which can be evaluated straightforwardly via the map (\ref{E}). Then the
reduced density matrices of interest are $\rho _{I}^{AB}(t)=$Tr$_{CD}\rho
_{I}^{ABCD}(t),$ $\rho _{I}^{CD}(t)=$Tr$_{AB}\rho _{I}^{ABCD}(t),$ $\rho
_{I}^{AC}(t)=$Tr$_{BD}\rho _{I}^{ABCD}(t)$ and $\rho _{I}^{BD}(t)=$Tr$%
_{AC}\rho _{I}^{ABCD}(t).$ All of $\rho _{I}^{AB}(t),$ $\rho _{I}^{CD}(t),$ $%
\rho _{I}^{AC}(t)$ and $\rho _{I}^{BD}(t)$ have the $X$-form so the
corresponding concurrences are determined by Eq. (\ref{CX}). In the
following we study the time dependence of these concurrences for the fields
in cavities $a$ and $b$ being initially in the vacuum state, the Fock state
with a non-zero photon number or the general thermal state, respectively.

\begin{figure}[tbp]
\centerline{\scalebox{0.8}{\includegraphics{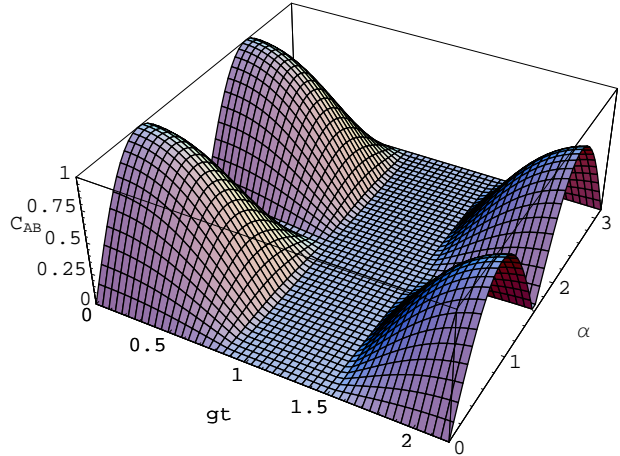}}}
\caption{The concurrence $C_{AB}\equiv C^{AB}_{I}(t)$ as functions of
rescaled time $gt$ and $\alpha $ for initially both cavity fields are in the
vacuum state and both atom-pairs $AB$ and $CD$ are in the $\left| \psi
(0)\right\rangle $(\ref{b1}) type state in the DTCM.}
\label{fig2}
\end{figure}

\begin{figure}[tbp]
\centerline{\scalebox{0.8}{\includegraphics{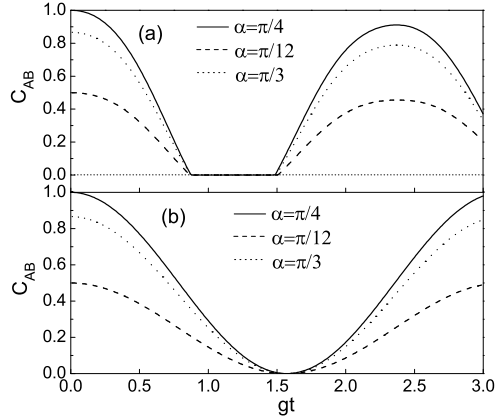}}}
\caption{The concurrence $C_{AB}\equiv C^{AB}_{I}(t)$ as a function of $gt$
for various values of $\alpha $ for the same initial preparation of cavities
and atoms as in Fig. 2 in (a) the DTCM and (b) the DJCM.}
\label{fig3}
\end{figure}

In FIG.2 we plot $C_{I}^{AB}$ (the same for $C_{I}^{BD}$ due to symmetry) as
functions of rescaled time $gt$ and $\alpha $ for the initial cavity fields
being in the vacuum state. From FIG. 2 it is transparent that $C_{I}^{AB}$
vanishes after a finite time of evolution and remains zero for some period
of time before increasing again. This dynamics holds in the whole range of $%
\alpha .$ A comparison between the DTCM and the DJCM [6] for the
same initial preparation of the cavities and atoms is shown in FIG.
3. Within the first cycle of evolution, in the DJCM (see FIG. 3b)
$C_{I}^{AB}$
vanishes at the moment $t_{0}=\pi /(2g)$ and grows up again right after $%
t_{0},$ while in the DTCM (see FIG. 3a) $C_{I}^{AB}=0$ at a time shorter
than $t_{0}$ and remains so for some time before reviving. This indicates
that for one and the same empty cavity fields, $\left| \psi (0)\right\rangle
$ type initial state of atoms does not undergo ESD in the DJCM but it does
in the DTCM. Therefore, the atomic entanglement dynamics is model-dependent
apart from the entanglement type itself. The physical interpretation behind
such a clear distinction in the dynamical behaviors between the two models
can be thought of as follows. If the cavities are empty, atoms in the ground
state $\left| 0\right\rangle $ remain unchanged and only atoms in the
excited state $\left| 1\right\rangle $ can interact with the cavity fields.
Denoting by $N_{\left| 1\right\rangle }$ the number of atoms that may be
populated in state $\left| 1\right\rangle ,$ the system-environment
interaction can be classified into two regimes, ``strong'' and ``weak''
interaction regimes, depending on relative magnitudes of $P_{\geq }$ and $%
P_{<},$ where $P_{\geq }$ $(P_{<})$ is the probability that
$N_{\left| 1\right\rangle }\geq N_{c}$ $(N_{\left| 1\right\rangle
}<N_{c})$ with $N_{c}$ the number of cavities. In the DTCM
considered here and the DJCM considered in [6,7] it is clear that
$N_{c}=2.$ We define the following convention: the strong
interaction regime corresponds to $P_{\geq }>P_{<},$ while $P_{\geq
}\leq P_{<}$ implies the weak interaction regime. In the DJCM the
total system state of two atoms $A,B$ and two cavities $a,b$ at
$t=0$ reads
\begin{equation}
\left| \psi (0)\right\rangle _{AB}\left| 00\right\rangle _{ab}=\cos \alpha
|10\rangle _{Aa}|00\rangle _{Bb}+\sin \alpha |00\rangle _{Aa}|10\rangle
_{Bb},  \label{djcmI}
\end{equation}
whereas in the DTCM the total system state of four atoms $A,B,C,D$ and two
cavities $a,b$ at $t=0$ reads
\begin{eqnarray}
\left| \psi (0)\right\rangle _{AB}\left| \psi (0)\right\rangle _{CD}\left|
00\right\rangle _{ab} &=&\cos ^{2}\alpha \left| 110\right\rangle
_{ACa}\left| 000\right\rangle _{BDb}+\cos \alpha \sin \alpha \left|
100\right\rangle _{ACa}\left| 010\right\rangle _{BDb}  \nonumber \\
&&+\sin \alpha \cos \alpha \left| 010\right\rangle _{ACa}\left|
100\right\rangle _{BDb}+\sin ^{2}\alpha \left| 000\right\rangle _{ACa}\left|
110\right\rangle _{BDb}.  \label{dtcmI}
\end{eqnarray}
From Eq. (\ref{djcmI}) it follows that there is always only one atom
(namely, either atom $A$ in the first term or atom $B$ in the second term)
being in state $\left| 1\right\rangle $ regardless of the value of $\alpha .$
That is, $P_{<}=1<P_{\geq }=0,$ resulting in the weak interaction regime in
the DJCM for the whole range of $\alpha .$ However, what is followed from
Eq. (\ref{dtcmI}) is that for any value of $\alpha $ there are always two
atoms (namely, either atoms $A$ and $C$ in the first term or atoms $A$ and $%
D $ in the second term or atoms $C$ and $B$ in the third term or atoms $B$
and $D$ in the fourth term) being in state $\left| 1\right\rangle .$ That
is, $P_{\geq }=1>P_{<}=0,$ resulting in the strong interaction regime in the
DTCM regardless of the value of $\alpha .$ Therefore, it can be said that,
when the cavities are initially prepared in the vacuum state, $\left| \psi
(0)\right\rangle $ type initial state of atoms exhibits ESD in the strong
interaction regime (i.e., in the DTCM) but it does not in the weak
interaction regime (i.e., in the DJCM), independent of the parameter $\alpha
.$

\begin{figure}[tbp]
\centerline{\scalebox{0.8}{\includegraphics{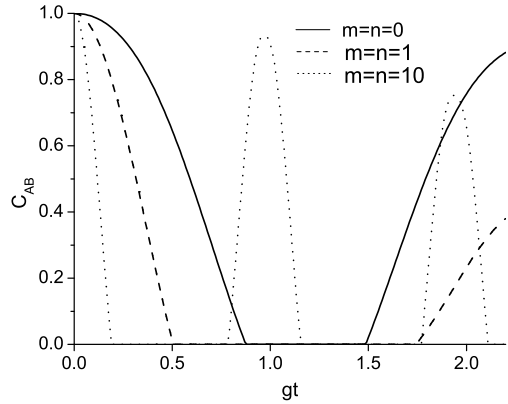}}}
\caption{The concurrence $C_{AB}\equiv C^{AB}_{I}(t)$ as a function of $gt$
for $\alpha =\pi/4$ for initially the cavity fields are in different Fock
states $\left| mn\right\rangle_{ab} $ and atom-pairs $AB$ and $CD$ are in
the $\left| \psi (0)\right\rangle $ (\ref{b1}) type state in the DTCM.}
\label{fig4}
\end{figure}

\begin{figure}[tbp]
\centerline{\scalebox{0.8}{\includegraphics{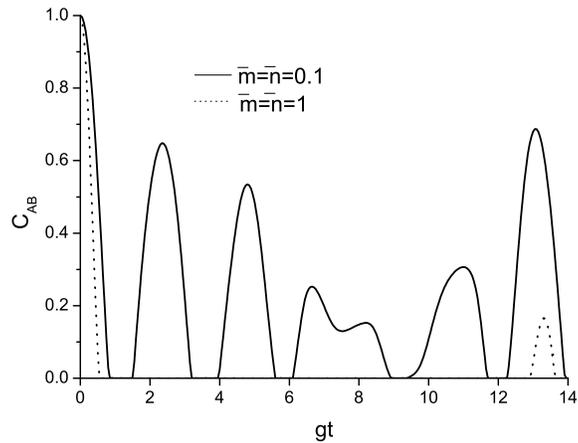}}}
\caption{The concurrence $C_{AB}\equiv C^{AB}_{I}(t)$ as a function of $gt$
for $\alpha =\pi/4 $ for initially cavity fields are in the thermal state
with different mean photon numbers $\overline{m}$, $\overline{n}$ and
atom-pairs $AB$ and $CD$ are in the $\left| \psi (0)\right\rangle $ (\ref{b1}%
) type state in the DTCM.}
\label{fig5}
\end{figure}

The case when the initial cavity fields are in a Fock state with a certain
nonzero photon number is illustrated in FIG. 4. In this case not only atoms
in state $\left| 1\right\rangle $ but also atoms in state $\left|
0\right\rangle ,$ i.e., all the present atoms, can interact with the cavity
fields so that the interaction regime is always strong resulting in ESD for
whatever values of $\alpha .$ A remarkable feature is that $C_{I}^{AB}$
decays quicker and reaches zero in a shorter time for a larger initial
number of photons in the cavities. The underlying physics for that feature
is the intensification of the system-environment effective interaction with
the increase of photon number contained in the cavities.

Figure 5 plots the evolution of $C_{I}^{AB}$ for the cavity fields being
initially in the thermal state. The entanglement dynamics looks chaotic due
to the nature of the thermal fields. As can be seen from FIG. 5, the larger
the mean photon number (corresponding to the higher temperature) the shorter
the death time of $C_{I}^{AB}$ and the longer its revival time.

\begin{figure}[tbp]
\centerline{\scalebox{0.8}{\includegraphics{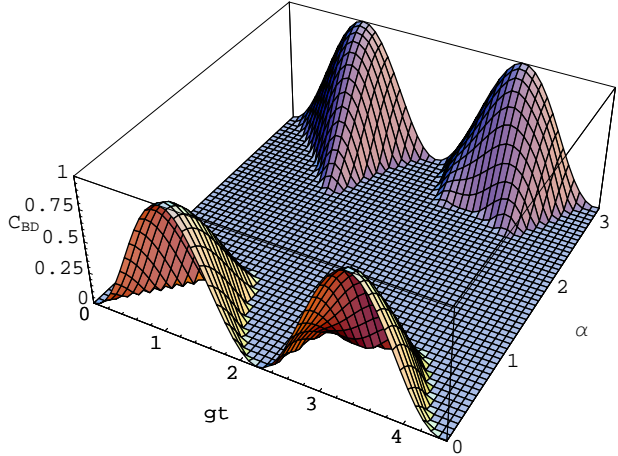}}}
\caption{The concurrence $C_{BD}\equiv C^{BD}_{I}(t)$ as functions of $gt$
and $\alpha $ for initially the cavity fields are in the Fock state $\left|
11\right\rangle_{ab}$ and both atom-pairs $AB$ and $CD$ are in the state (%
\ref{b1}) in the DTCM.}
\label{fig6}
\end{figure}

\begin{figure}[tbp]
\centerline{\scalebox{0.8}{\includegraphics{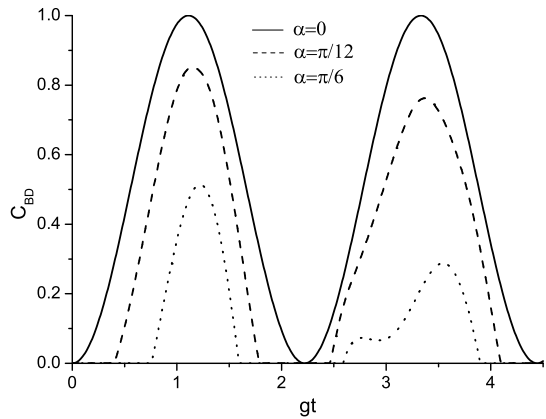}}}
\caption{The concurrence $C_{BD}\equiv C^{BD}_{I}(t)$ as a function of $gt$
for various values of $\alpha $ with the same initial preparation of cavity
fields and atom-pairs as in Fig. 6 in the DTCM.}
\label{fig7}
\end{figure}

\begin{figure}[tbp]
\centerline{\scalebox{0.8}{\includegraphics{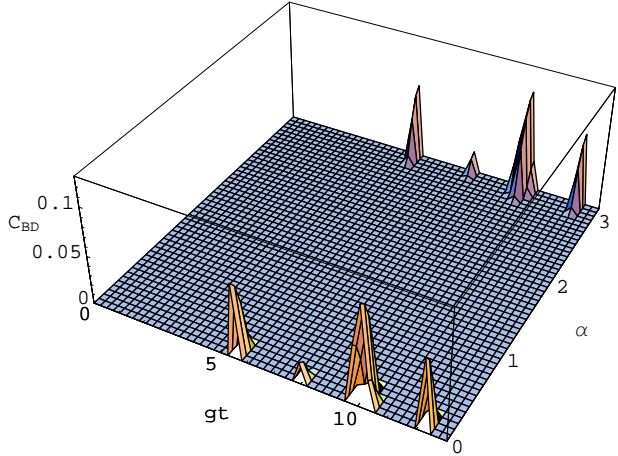}}} \caption{The
concurrence $C_{BD}\equiv C^{BD}_{I}(t)$ as functions of $gt$ and
$\alpha $ for initially the cavity fields are in the thermal state
with
the mean photon numbers $\overline{m}=\overline{n}=1$ and both atom-pairs $%
AB $ and $CD$ are in the state (\ref{b1}) in the DTCM.} \label{fig8}
\end{figure}

At this point let us study the dynamics of the two atoms that are
located in one and the same cavity. These are atoms $A$ and $C$ in
cavity $a$ and atoms $B$ and $D$ in cavity $b.$ Such atoms in the
same cavity are absolutely uncorrelated at the beginning and also
there are no direct interactions between them during the entire
course of evolution, in accordance with the problem Hamiltonians
(\ref{H1}) and (\ref{H2}). However, an effective (indirect)
atom-atom interaction is induced for $t>0$ thanks to the coupling of
both atoms with a common environment. Such an effective atom-atom
interaction could nontrivially affect their global behaviors. In
fact, as investigated in Ref. [17], if the initial atoms are
prepared either in state $\left| 01\right\rangle $ or $\left|
10\right\rangle $ $(\left| 11\right\rangle )$, then they always get
entangled with each other (remain unentangled) regardless of the
nature of the cavity fields. But, if the atomic initial state is
$\left| 00\right\rangle $, then the field in the vacuum state leaves
the atoms unentangled and the field in a Fock state with a non-zero
photon number or thermal state can entangle them. Here, in the DTCM,
at variance with the situation considered in Ref. [17], at $t=0$ the
atoms in a cavity, though being independent of each other, are
entangled
with other atoms in another cavity. That is, we have at $t=0$ in cavity $a$ $%
(b)$ a mixed state $\rho _{I}^{AC}(0)=$Tr$_{BD}\rho
_{I}^{ABCD}(0)=\sum_{i,j=0}^{1}\alpha _{i}^{2}\alpha _{j}^{2}\left|
i,j\right\rangle _{ACAC}\left\langle i,j\right| $ $(\rho _{I}^{BD}(0)=$Tr$%
_{AC}\rho _{I}^{ABCD}(0)=\sum_{i,j=0}^{1}\alpha _{i}^{2}\alpha
_{j}^{2}\left| i\oplus 1,j\oplus 1\right\rangle _{BDBD}\left\langle
i\oplus 1,j\oplus 1\right| ),$ instead of a pure state as in Ref.
[17]. Figure 6 plots the concurrence $C_{I}^{BD}$ as functions of
$gt$ and $\alpha $ with the initial fields in both cavities
containing just one photon. This
figure shows that the entanglement dynamics of the atoms is sensitive to $%
\alpha ,$ as it should be. For example, in the region of $\alpha \in
[0,0.29\pi ]$ atoms $B$ and $D$ can get entangled, but for $\alpha $ around $%
\pi /2$ no entanglement is generated through the whole evolution. These
results are in full agreement with those reported in Ref. [17] where $%
\alpha =0$ (i.e., $\rho _{I}^{BD}(0)=\left| 00\right\rangle
_{BDBD}\left\langle 00\right| )$ and $\alpha =\pi /2$ (i.e., $\rho
_{I}^{BD}(0)=\left| 11\right\rangle _{BDBD}\left\langle 11\right| )$
are concerned. To get more insight into the effect of $\alpha $ on
atomic entanglement generation we show in FIG. 7 a 2D plot of
$C_{I}^{BD}$ as a function of $gt$ with the initial cavity fields in
the Fock states $\left| 1,1\right\rangle _{ab}$ for various values
of $\alpha .$ When $\alpha =0$ (i.e., $\rho _{I}^{BD}(0)=\left|
00\right\rangle _{BDBD}\left\langle 00\right| ),$ the entanglement
of $B$ and $D$ emerges immediately from $t=0.$ Nevertheless, when
$\alpha >0$ the atoms remain unentangled for some initial period of
time and suddenly become entangled at some later time. The larger
the value of $\alpha $ the longer the delay time of entanglement
generation. Such phenomena of delayed entanglement during the time
evolution can be called ``entanglement sudden birth'' (ESB) [22].
The effect of thermal fields on inducing entanglement between atoms
$B$ and $D$ is drawn in FIG. 8 with the cavity mean photon numbers
$\overline{m}=\overline{n}=1,$ which agrees well with the result in
Ref. [17] for $\alpha =0.$ Since the thermal state is a weighted
mixture of Fock states (see Eq. (\ref{roE})), it is a chaotic state
with minimum information and so its effect is generally
irregular. In comparison with the case of ``corresponding'' Fock states $%
\left| 1,1\right\rangle _{ab}$ one sees that the region of $\alpha $
allowing entanglement of atoms is much shrunk and the amount of generated
entanglement is very small. The plots of $C_{I}^{AC}$ can be obtained from
those of $C_{I}^{BD}$ by making a change $\alpha \rightarrow \alpha +\pi /2.$

\vskip 0.8cm \noindent {\bf 3.2 $\left| \varphi(0)\right\rangle $
type initial state for atom-pairs $AB$ and $CD$}\vskip 0.5cm

We next consider the case when both atom-pairs $AB$ and $CD$ are initially
prepared in state (\ref{b2}). In accordance with Eq. (\ref{rABCD}) the
reduced density matrix of the atomic subsystem at any time $t$ is
\begin{equation}
\rho _{II}^{ABCD}(t)=\sum_{i,j,k,l=0}^{1}\alpha _{i}\alpha _{j}\alpha
_{k}\alpha _{l}\mathcal{E}_{AC}^{a}\left( \left| i,k\right\rangle
_{ACAC}\left\langle j,l\right| \right) \otimes \mathcal{E}_{BD}^{b}\left(
\left| i,k\right\rangle _{BDBD}\left\langle j,l\right| \right) .
\end{equation}

In FIG.9 we plot $C_{II}^{AB}$ (the same for $C_{II}^{BD}$ due to symmetry)
versus $gt$ and $\alpha $ for the initial empty cavity fields. It is visual
from this figure that ESD occurs but not in the whole range of $\alpha ,$ in
clear contrast with the case shown in FIG. 2 when both the atom-pairs $AB$
and $CD$ are initially prepared in state (\ref{b1}). To derive the
constraint on $\alpha $ that triggers ESD let us look at the total system
state at $t=0:$
\begin{eqnarray}
\left| \varphi (0)\right\rangle _{AB}\left| \varphi (0)\right\rangle
_{CD}\left| 00\right\rangle _{ab} &=&\cos ^{2}\alpha \left| 110\right\rangle
_{ACa}\left| 110\right\rangle _{BDb}+\cos \alpha \sin \alpha \left|
100\right\rangle _{ACa}\left| 100\right\rangle _{BDb}  \nonumber \\
&&+\sin \alpha \cos \alpha \left| 010\right\rangle _{ACa}\left|
010\right\rangle _{BDb}+\sin ^{2}\alpha \left| 000\right\rangle _{ACa}\left|
000\right\rangle _{BDb}.  \label{dtcmII}
\end{eqnarray}
Obviously, the probability that all the four atoms are in state $\left|
1\right\rangle $ is $\cos ^{4}\alpha ,$ the probability that only two atoms
(namely, either atoms $A$ and $B$ or atoms $C$ and $D)$ are in state $\left|
1\right\rangle $ is $2\cos ^{2}\alpha \sin ^{2}\alpha $ and the probability
that none of the atoms are in state $\left| 1\right\rangle $ (i.e., all the
atoms are in state $\left| 0\right\rangle )$ is $\sin ^{4}\alpha .$ That is,
$P_{\geq }=\cos ^{4}\alpha +2\cos ^{2}\alpha \sin ^{2}\alpha $ and $%
P_{<}=\sin ^{4}\alpha .$ As mentioned in the previous subsection, the
condition for the occurrence of ESD is that the interaction regime is
strong, i.e., $P_{\geq }>P_{<}.$ So, the values of $\alpha $ for which ESD
occurs should satisfy the constraint
\begin{equation}
\sin ^{2}\alpha <\frac{1}{\sqrt{2}}.  \label{t}
\end{equation}
Noticeably, this constraint is not coincident with that one in the DJCM for
which the initial total system state reads
\begin{equation}
\left| \varphi (0)\right\rangle _{AB}\left| 00\right\rangle _{ab}=\cos
\alpha |10\rangle _{Aa}|10\rangle _{Bb}+\sin \alpha |00\rangle
_{Aa}|00\rangle _{Bb}.  \label{djcmII}
\end{equation}
As followed from Eq. (\ref{djcmII}), the probability that the two atoms are
in state $\left| 1\right\rangle $ is $\cos ^{2}\alpha $ and the probability
that none of the atoms are in state $\left| 1\right\rangle $ is $\sin
^{2}\alpha .$ That is, $P_{\geq }=\cos ^{2}\alpha ,$ $P_{<}=\sin ^{2}\alpha $
and thus the values of $\alpha ,$ for which the system-environment
interaction regime is strong (i.e., ESD occurs) in the DJCM, satisfy the
constraint
\begin{equation}
\sin ^{2}\alpha <\frac{1}{2}.  \label{j}
\end{equation}
The constraints (\ref{t}) and (\ref{j}) imply that the $\alpha $-parameter
domain in which the atoms suffer from ESD is wider in the DTCM than in the
DJCM.

\begin{figure}[tbp]
\centerline{\scalebox{0.8}{\includegraphics{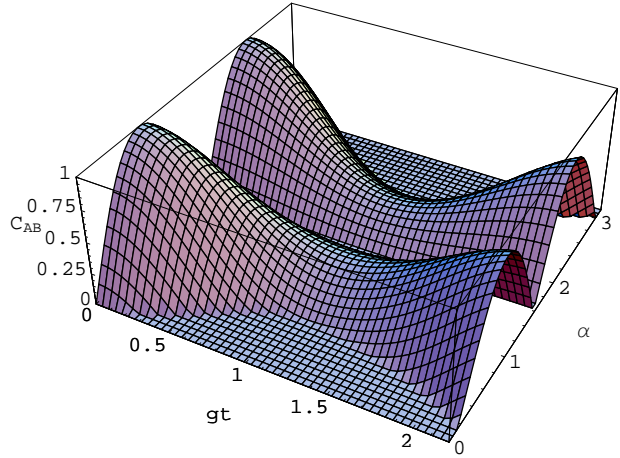}}}
\caption{The concurrence $C_{AB}\equiv C^{AB}_{II}(t)$ as functions of $gt$
and $\alpha $ for initially both cavity fields are in the vacuum state and
both atom-pairs $AB$ and $CD$ are in the $\left| \varphi (0)\right\rangle $ (%
\ref{b2}) type state in the DTCM.}
\label{fig9}
\end{figure}

\begin{figure}[tbp]
\centerline{\scalebox{0.8}{\includegraphics{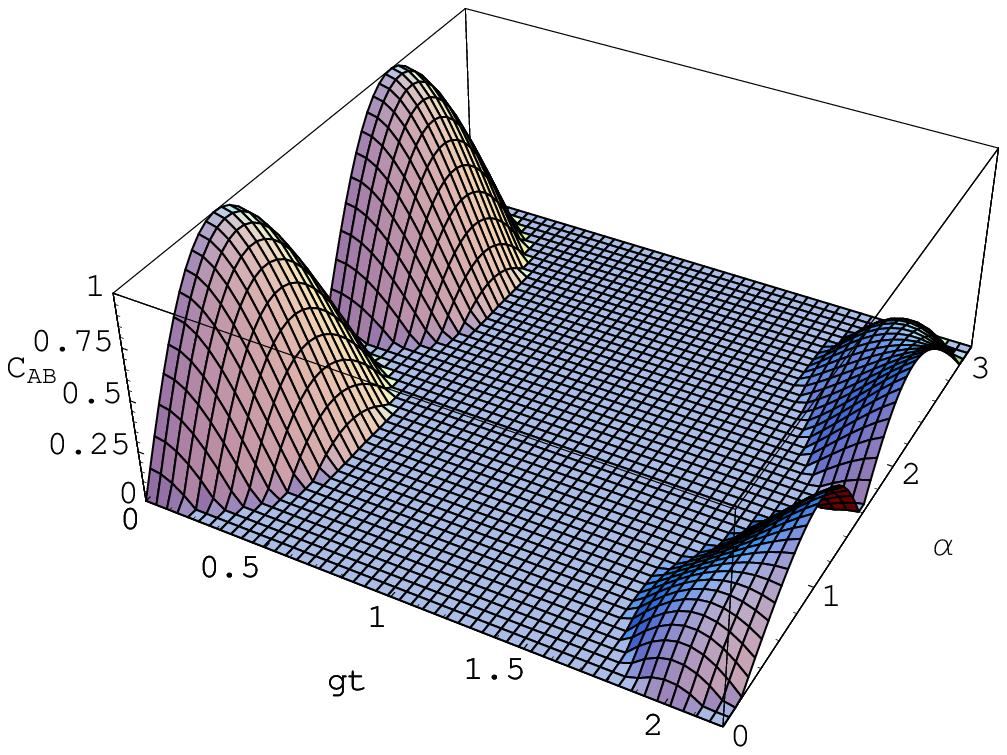}}}
\caption{The concurrence $C_{AB}\equiv C^{AB}_{II}(t)$ as functions of $gt$
and $\alpha$ for initially the cavity fields are in the Fock state $\left|
11\right\rangle_{ab} $ and both atom-pairs $AB$ and $CD$ are in the $\left|
\varphi (0)\right\rangle $ (\ref{b2}) type state in the DTCM.}
\label{fig10}
\end{figure}

\begin{figure}[tbp]
\centerline{\scalebox{0.8}{\includegraphics{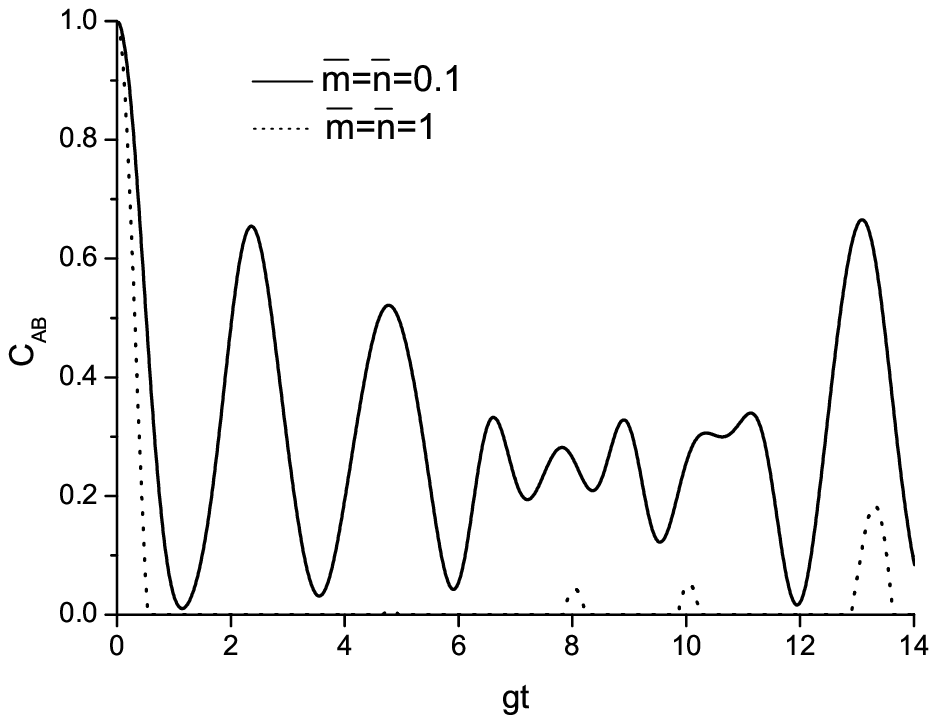}}}
\caption{The concurrence $C_{AB}\equiv C^{AB}_{II}(t)$ as a function of $gt$
for $\alpha =\pi/4 $ for initially the cavity fields are in the thermal
state with different mean photon numbers $\overline{m}$, $\overline{n}$ and
both atom-pairs $AB$ and $CD$ are in the $\left| \varphi (0)\right\rangle $ (%
\ref{b2}) type state in the DTCM.}
\label{fig11}
\end{figure}

The case for the initial cavity fields being in a Fock state $\left|
11\right\rangle_{ab} $ is plotted in FIG. 10. A remarkable feature as
compared with the vacuum fields case in FIG. 9 is that here ESD occurs in
the whole range of $\alpha .$ Again, the physical reason for this is that in
the presence of initial photons all the atoms are in interaction with the
cavity fields (i.e., not only atoms in state $\left| 1\right\rangle $ but
also those in state $\left| 0\right\rangle $ interact with the cavity
fields).

In FIG. 11 we plot $C_{II}^{AB}$ as a function of $gt$ for the initial
fields in a thermal state with different mean photon numbers for a given
value of $\alpha .$ Comparing FIG. 11 with FIG. 5 signals that with
relatively small mean photon numbers (e.g., $\overline{m}=\overline{n}=0.1$)
the signature of ESD is less pronounced for the case when the initial atoms
are prepared in state (\ref{b2}) than in state (\ref{b1}).

The entanglement generation dynamics of the atomic pairs $AC$ and $BD$ is
similar to the case considered in the preceding subsection and thus will not
be iterated here.

\vskip 0.8cm \noindent {\bf 4 Conclusion}\vskip 0.5cm

\noindent In conclusion, we have, by means of concurrence, studied
the entanglement dynamics of the DTCM motivated by certain realistic
quantum information processing. The system is composed of four
two-level atoms $A,B,C,D$ and two spatially separated single-mode
cavities $a,b$. Initially, atom-pairs $AB$ and $CD$ are prepared
either in Bell-like state $\left| \psi
(0)\right\rangle $ (\ref{b1}) or $\left| \varphi (0)\right\rangle $ (\ref{b2}%
), while both cavities are prepared either in the vacuum state, the Fock
state with non-zero photon numbers or the thermal sate. Independent atoms $%
A,C$ ($B,D$) that belong to different entangled atom-pairs are embedded in
one and the same cavity $a$ ($b$) and interact with it through the
Tavis-Cummings Hamiltonian.

For the vacuum fields the $\left| \psi (0)\right\rangle $ type
initial state of atom-pairs $AB$ and $CD$ displays ESD for the whole
value range of the parameter $\alpha $ which represents the initial
entanglement degree of $AB$ and $CD$. This result is in sharp
contrast with the DJCM for which ESD does not occur at all for
whatever values of $\alpha $ [6,7]. As for the $\left| \varphi
(0)\right\rangle $ type initial state of atom-pairs $AB$ and $CD$,
ESD only occur for the value of $\alpha $ such that $\sin ^{2}\alpha
<1/\sqrt{2}$, which is wider than that in the DJCM where ESD
occurs just for $\alpha $ such that $\sin ^{2}\alpha <1/2$ [6,7]%
. Physically, these results (i.e., the domain of $\alpha $ for which ESD
occurs) in both the DTCM and DJCM can be explained via the interaction
strength theory according to which ESD occurs (does not occur) in the strong
(weak) system-environment interaction regime. The interaction regime is
identified by the number of atoms that can have interaction with the
cavities, which is determined by the relative magnitudes of $P_{\geq }$ and $%
P_{<}$ defined in subsection 3.1. Remarkably, the interaction
strength theory turns out to apply also for the so-called triple
Jaynes-Cummings model [23] for GHZ-like atomic states as well as for
the case of multiple dissipative environments with multiqubit
GHZ-like atomic states [24,25].

We have shown that the non-vacuum environments of cavities have great
effects on the appearance of ESD for atoms. That is, when the cavity fields
are initially in the Fock state with a non-zero photon number or the general
thermal state, ESD always happens for atom-pairs $AB$ and $CD$ regardless of
the entanglement type they are prepared. Moreover, the more photon number in
the Fock state or the greater the mean photon number in the thermal state
the quicker the entanglement decay rate, i.e., the sooner the time of ESD
occurrence. In terms of the interaction strength theory, these properties
are explained by the physical fact that in the presence of nonzero (mean)
photon number the interaction regime is always strong because all the atoms
(i.e., not only those in the excited state as in the case of empty cavities)
can interact with the fields. Thus, the actual system-environment
interaction strength is now identified by the number of excitation which in
these cases is proportional to the total number of both atoms and photons.

We have also studied creation of entanglement between initially uncorrelated
atoms $A$ and $C$ in cavity $a$ ($B$ and $D$ in cavity $b).$ Compared to the
case of $\alpha =0$ considered in Ref. [17] here we showed that for $%
\alpha \neq 0$ there appears the so-called entanglement sudden
birth, i.e., the formation of atomic entanglement does not take
place at once as the system evolves but emerges suddenly at some
delayed time, which is dependent on the value of $\alpha .$ The DTCM
presented in this work could be extended to the general multiple
case where two groups of multipartite entangled atoms are
distributed in such a way that every two atoms from different group
are located in the same environment. In this way, we can study not
only the pairwise entanglement of atoms between any two\textbf{\
}nodes (cavities or local environments) via concurrence but also the
entanglement of any atomic bipartition by means of negativity. These
studies can reveal the degraded properties of various multipartite
entangled state and thus be useful for the large-scale quantum
information processing.\vskip 1cm

\noindent Z.X.M. and Y.J.X. are supported by National Natural
Science Foundation of China under Grant No. 10774088. N.B.A.
acknowledges support from a NAFOSTED project No. NCCB-2009 and from
the KIAS Scholar program.

\appendix

\section{The explicit expressions of $X_{ik,pq}(m,\tau )$}

The functions $X_{ik,pq}(m,\tau )$ appearing in Eq. (\ref{UACa}) for all
possible $i,k,p,q$ are given by
\begin{equation}
X_{11,00}(m,\tau )=\frac{m+1}{2m+3}[\cos (\sqrt{2(2m+3)}\tau )-1]+1,
\end{equation}
\begin{equation}
X_{11,10}(m,\tau )=X_{11,01}(m,\tau )=-i\sqrt{\frac{m+1}{2(2m+3)}}\sin (%
\sqrt{2(2m+3)}\tau ),
\end{equation}
\begin{equation}
X_{11,11}(m,\tau )=\frac{\sqrt{(m+1)(m+2)}}{2m+3}[\cos (\sqrt{2(2m+3)}\tau
)-1],
\end{equation}
\begin{equation}
X_{01,10}(m,\tau )=X_{10,01}(m,\tau )=-i\sqrt{\frac{m}{2(2m+1)}}\sin (\sqrt{%
2(2m+1)}\tau ),
\end{equation}
\begin{equation}
X_{01,00}(m,\tau )=X_{10,00}(m,\tau )=\frac{1}{2}[\cos (\sqrt{2(2m+1)}\tau
)+1],
\end{equation}
\begin{equation}
X_{01,11}(m,\tau )=X_{10,11}(m,\tau )=\frac{1}{2}[\cos (\sqrt{2(2m+1)}\tau
)-1],
\end{equation}
\begin{equation}
X_{01,01}(m,\tau )=X_{10,10}(m,\tau )=-i\sqrt{\frac{m+1}{2(2m+1)}}\sin (%
\sqrt{2(2m+1)}\tau ),
\end{equation}
\begin{equation}
X_{00,11}(m,\tau )=\frac{\sqrt{m(m-1)}}{2m-1}[\cos (\sqrt{2(2m-1)}\tau )-1],
\end{equation}
\begin{equation}
X_{00,01}(m,\tau )=X_{00,10}(m,\tau )=-i\sqrt{\frac{m}{2(2m-1)}}\sin (\sqrt{%
2(2m-1)}\tau )
\end{equation}
and
\begin{equation}
X_{00,00}(m,\tau )=\frac{m}{2m-1}[\cos (\sqrt{2(2m-1)}\tau )-1]+1.
\end{equation}

\section{The explicit expressions of $\mathcal{E}_{XY}^{c}\left( \left|
ik\right\rangle _{XYXY}\left\langle jl\right| \right) $}

The expressions of the map $\mathcal{E}_{XY}^{c}\left( \left|
ik\right\rangle _{XYXY}\left\langle jl\right| \right) ,$ with $XYc=ACa$ or $%
BDb,$ appearing in Eq. (\ref{E}) for all possible $i,k,j,l$ are given by
\begin{eqnarray}
\mathcal{E}_{XY}^{c}\left( |00\rangle _{XYXY}\langle 00|\right)
&=&\sum_{m=0}^{\infty }P_{m}^{c}\left[ |X_{00,11}(m,\tau )|^{2}|11\rangle
_{XYXY}\langle 11|\right.   \nonumber \\
&&+|X_{00,10}(m,\tau )|^{2}|10\rangle _{XYXY}\langle 10|  \nonumber \\
&&+X_{00,10}(m,\tau )X_{00,01}^{*}(m,\tau )|10\rangle _{XYXY}\langle 01|
\nonumber \\
&&+X_{00,01}(m,\tau )X_{00,10}^{*}(m,\tau )|01\rangle _{XYXY}\langle 10|
\nonumber \\
&&+|X_{00,01}(m,\tau )|^{2}|01\rangle _{XYXY}\langle 01|  \nonumber \\
&&\left. +|X_{00,00}(m,\tau )|^{2}|00\rangle _{XYXY}\langle 00|\right] ,
\label{e0000}
\end{eqnarray}
\begin{eqnarray}
\mathcal{E}_{XY}^{c}\left( |01\rangle _{XYXY}\langle 00|\right)  &=&\mathcal{%
E}_{XY}^{c}\left( |00\rangle _{XYXY}\langle 01|\right) ^{*}  \nonumber \\
&=&\sum_{m=0}^{\infty }P_{m}^{c}\left[ X_{01,10}(m,\tau
)X_{00,01}^{*}(m,\tau )|11\rangle _{XYXY}\left( \langle 01|+\langle
10|\right) |\right.   \nonumber \\
&&+X_{01,11}(m,\tau )X_{00,00}^{*}(m,\tau )|10\rangle _{XYXY}\langle 00|
\nonumber \\
&&\left. +X_{01,00}(m,\tau )X_{00,00}^{*}(m,\tau )|01\rangle _{XYXY}\langle
00|\right] ,  \label{e0100}
\end{eqnarray}
\begin{eqnarray}
\mathcal{E}_{XY}^{c}\left( |10\rangle _{XYXY}\langle 00|\right)  &=&\mathcal{%
E}_{XY}^{c}\left( |00\rangle _{XYXY}\langle 10|\right) ^{*}  \nonumber \\
&=&\sum_{m=0}^{\infty }P_{m}^{c}\left[ X_{10,01}(m,\tau
)X_{00,10}^{*}(m,\tau )|11\rangle _{XYXY}\langle 10|\right.   \nonumber \\
&&+X_{10,01}(m,\tau )X_{00,01}^{*}(m,\tau )|11\rangle _{XYXY}\langle 01|
\nonumber \\
&&+X_{10,00}(m,\tau )X_{00,00}^{*}(m,\tau )|10\rangle _{XYXY}\langle 00|
\nonumber \\
&&\left. +X_{10,11}(m,\tau )X_{00,00}^{*}(m,\tau )|01\rangle _{XYXY}\langle
00|\right] ,  \label{e1000}
\end{eqnarray}
\begin{eqnarray}
\mathcal{E}_{XY}^{c}\left( |11\rangle _{XYXY}\langle 00|\right)  &=&\mathcal{%
E}_{XY}^{c}\left( |00\rangle _{XYXY}\langle 11|\right) ^{*}  \nonumber \\
&=&\sum_{m=0}^{\infty }P_{m}^{c}X_{11,00}(m,\tau )X_{00,00}^{*}(m,\tau
)|11\rangle _{XYXY}\langle 00|,  \label{e1100}
\end{eqnarray}
\begin{eqnarray}
\mathcal{E}_{XY}^{c}\left( |01\rangle _{XYXY}\langle 01|\right)
&=&\sum_{m=0}^{\infty }P_{m}^{c}\left[ |X_{01,10}(m,\tau )|^{2}|11\rangle
_{XYXY}\langle 11|\right.   \nonumber \\
&&+X_{01,11}(m,\tau )X_{01,00}^{*}(m,\tau )|10\rangle _{XYXY}\langle 01|
\nonumber \\
&&+|X_{01,11}(m,\tau )|^{2}|10\rangle _{XYXY}\langle 00|  \nonumber \\
&&+X_{01,00}(m,\tau )X_{01,11}^{*}(m,\tau )|01\rangle _{XYXY}\langle 10|
\nonumber \\
&&+|X_{01,00}(m,\tau )|^{2}|01\rangle _{XYXY}\langle 01|  \nonumber \\
&&\left. +|X_{01,01}(m,\tau )|^{2}|00\rangle _{XYXY}\langle 00|\right] ,
\label{e0101}
\end{eqnarray}
\begin{eqnarray}
\mathcal{E}_{XY}^{c}\left( |10\rangle _{XYXY}\langle 01|\right)  &=&\mathcal{%
E}_{XY}^{c}\left( |01\rangle _{XYXY}\langle 10|\right) ^{*}  \nonumber \\
&=&\sum_{m=0}^{\infty }P_{m}^{c}\left[ |X_{10,01}(m,\tau )|^{2}|11\rangle
_{XYXY}\langle 11|\right.   \nonumber \\
&&+X_{10,00}(m,\tau )X_{01,11}^{*}(m,\tau )|10\rangle _{XYXY}\langle 10|
\nonumber \\
&&+|X_{10,00}(m,\tau )|^{2}|10\rangle _{XYXY}\langle 01|  \nonumber \\
&&+|X_{10,11}(m,\tau )|^{2}|01\rangle _{XYXY}\langle 10|  \nonumber \\
&&+X_{10,11}(m,\tau )X_{01,00}^{*}(m,\tau )|01\rangle _{XYXY}\langle 01|
\nonumber \\
&&\left. +|X_{10,10}(m,\tau )|^{2}|00\rangle _{XYXY}\langle 00|\right] ,
\label{e1001}
\end{eqnarray}
\begin{eqnarray}
\mathcal{E}_{XY}^{c}\left( |11\rangle _{XYXY}\langle 01|\right)  &=&\mathcal{%
E}_{XY}^{c}\left( |01\rangle _{XYXY}\langle 11|\right) ^{*}  \nonumber \\
&=&\sum_{m=0}^{\infty }P_{m}^{c}\left[ X_{11,00}(m,\tau
)X_{01,11}^{*}(m,\tau )|11\rangle _{XYXY}\langle 10|\right.   \nonumber \\
&&+X_{11,00}(m,\tau )X_{01,00}^{*}(m,\tau )|11\rangle _{XYXY}\langle 01|
\nonumber \\
&&+X_{11,01}(m,\tau )X_{01,01}^{*}(m,\tau )|10\rangle _{XYXY}\langle 00|
\nonumber \\
&&\left. +X_{11,10}(m,\tau )X_{01,01}^{*}(m,\tau )|01\rangle _{XYXY}\langle
00|\right] ,
\end{eqnarray}
\begin{eqnarray}
\mathcal{E}_{XY}^{c}\left( |10\rangle _{XYXY}\langle 10|\right)
&=&\sum_{m=0}^{\infty }P_{m}^{c}\left[ |X_{10,01}(m,\tau )|^{2}|11\rangle
_{XYXY}\langle 11|\right.   \nonumber \\
&&+|X_{10,00}(m,\tau )|^{2}|10\rangle _{XYXY}\langle 10|  \nonumber \\
&&+X_{10,00}(m,\tau )X_{10,11}^{*}(m,\tau )|10\rangle _{XYXY}\langle 01|
\nonumber \\
&&+X_{10,11}(m,\tau )X_{10,00}^{*}(m,\tau )|01\rangle _{XYXY}\langle 10|
\nonumber \\
&&+|X_{10,11}(m,\tau )|^{2}|01\rangle _{XYXY}\langle 01|+  \nonumber \\
&&\left. |X_{10,10}(m,\tau )|^{2}|00\rangle _{XYXY}\langle 00|\right] ,
\label{e1010}
\end{eqnarray}
\begin{eqnarray}
\mathcal{E}_{XY}^{c}\left( |11\rangle _{XYXY}\langle 10|\right)  &=&\mathcal{%
E}_{XY}^{c}\left( |10\rangle _{XYXY}\langle 11|\right) ^{*}  \nonumber \\
&=&\sum_{m=0}^{\infty }P_{m}^{c}\left[ X_{11,00}(m,\tau
)X_{10,00}^{*}(m,\tau )|11\rangle _{XYXY}\langle 10|\right.   \nonumber \\
&&+X_{11,00}(m,\tau )X_{10,11}^{*}(m,\tau )|11\rangle _{XYXY}\langle 01|
\nonumber \\
&&+X_{11,01}(m,\tau )X_{10,10}^{*}(m,\tau )|10\rangle _{XYXY}\langle 00|
\nonumber \\
&&+\left. X_{11,10}(m,\tau )X_{10,10}^{*}(m,\tau )|01\rangle _{XYXY}\langle
00|\right]   \label{e1110}
\end{eqnarray}
and
\begin{eqnarray}
\mathcal{E}_{XY}^{c}\left( |11\rangle _{XYXY}\langle 11|\right)
&=&\sum_{m=0}^{\infty }P_{m}^{c}\left[ |X_{11,00}(m,\tau )|^{2}|11\rangle
_{XYXY}\langle 11|\right.   \nonumber \\
&&+|X_{11,01}(m,\tau )|^{2}\left( |10\rangle +|01\rangle \right)
_{XYXY}\left( \langle 10|+\langle 01|\right)   \nonumber \\
&&\left. +|X_{11,11}(m,\tau )|^{2}|00\rangle _{XYXY}\langle 00|\right] .
\label{e1111}
\end{eqnarray}

\vskip 0.7cm \noindent {\bf References} \vskip 0.5cm

\noindent1. M. A. Nielsen and I. L. Chuang, Quantum Computation and
Quantum Information (Cambridge University Press. Cmabridge, 2000)

\noindent2. C. H. Bennett et al., Phys. Rev. Lett. \textbf{70}, 1895
(1993)

\noindent3. A. K. Pati, Phys. Rev. A \textbf{63}, 014302 (2000)

\noindent4. F. G. Deng, G. L. Long and X. S. Liu, Phys. Rev. A
\textbf{\ 68,} 042317 (2003)

\noindent5. M. Y\"{o}na\c{c}, T. Yu and J. H. Eberly, J. Phys. B
\textbf{39}, S621 (2006)

\noindent6. M. Y\"{o}na\c{c}, T. Yu T and J. H. Eberly, J. Phys. B
\textbf{40}, S45 (2007)

\noindent7. I. Sainz and G. Bj\"{o}rk, Phys. Rev. A \textbf{76},
042313 (2007)

\noindent8. J. G. Oliveira, R. Rossi, and M. C. Nemes, Phys. Rev. A
\textbf{78}, 044301 (2008)

\noindent9. D. Cavalcanti, et al., Phys. Rev. A \textbf{74}, 042328
(2006)

\noindent10. Z. X. Man, Y. J. Xia, and Nguyen Ba An, J. Phys. B \textbf{%
41}, 085503 (2008)

\noindent11. J. H. Cole, e-prient arXiv:quant-ph/0809.1746v1

\noindent12. S. Chan, M. D. Reid, and Z. Ficek, e-prient
arXiv:quant-ph/0810.3050v1

\noindent13. T. Yu and J. H. Eberly, Phys. Rev. Lett. \textbf{97},
140403 (2006);

J. H. Eberly and T. Yu, Science \textbf{316}, 555 (2007);

B. Bellomo, R. Lo Francl and G. Compagno, Phys. Rev. Lett.
\textbf{99}, 160502 (2007)

\noindent14. Almeida M P, \textit{et al.,} Science \textbf{316}, 579
(2007)

\noindent15. J. Laurat, K. S. Choi, H. Deng, C. W. Chou, H. J.
Kimble, Phys. Rev. Lett. \textbf{99}, 180504 (2007)

\noindent16. M. Tavis and F. W. Cunnings, Phys. Rev. \textbf{170},
379 (1968)

\noindent17. M. S. Kim, J. Lee, D. Ahn, and P. L. Knight, Phys. Rev.
A \textbf{65}, 040101(R) (2002)

\noindent18. H. T. Cui, K. Li, and X. X. Yi, Phys. Lett. A \textbf{\ 365}%
, 44 (2007)

\noindent19. W. K. Wootters, Phys. Rev. Lett. \textbf{80}, 2245
(1998)

\noindent20. G. Vidal and R. F. Werner, Phys. Rev. A \textbf{65},
032314(2002);

K. Audenaert, M. B. Plenio, and J. Eisert, Phys. Rev. Lett.
\textbf{90}, 027901 (2003)

\noindent21. T. Yu and J. H. Eberly, Quantum Inf. Comput.
\textbf{7}, 459 (2007)

\noindent22. C. E. L\'{o}pez, \textit{et al.,}, Phys. Rev. Lett.
\textbf{101}, 080503 (2008);

M. Abdel-Aty, T. Yu, J. Phys. B \textbf{41}, 235503 (2008);

Z. Ficek and R. Tanas, Phys. Rev. A \textbf{77}, 054301 (2008)

\noindent23. Z. X. Man, Y. J. Xia, and Nguyen Ba An, J. Phys. B \textbf{41%
}, 155501 (2008)

\noindent24. L. Aolita, R. Chaves, D. Cavalcanti, A. Ac\'{i}n, and
L. Davidovich, Phys. Rev. Lett. \textbf{100}, 080501 (2008)

\noindent25. Z. X. Man, Y. J. Xia, and Nguyen Ba An, Phys. Rev. A
\textbf{78}, 064301 (2008)

\end{document}